\documentstyle[12pt]{article}
\parskip=\medskipamount

\newcommand {\cD}{{\cal D}}

\newcommand {\cU}{{\cal U}}

\def\d{\delta}

\def\q{\theta}

\def\t{\tau}

\newcommand{\be}{\begin{equation}}
\newcommand{\ee}{\end{equation}}
\newcommand{\bea}{\begin{eqnarray}}
\newcommand{\eea}{\end{eqnarray}}
\newcommand{\non}{\nonumber}
\begin{document}

\begin{titlepage}
\thispagestyle{empty}

\begin{flushleft}
IASSNS-HEP-9732T\\
UPR-745T\\
ITP-UH-12/97 \\
hep-th/9704214 \\
\end{flushleft}

\begin{center}
{\large\bf The Background Field Method for N = 2 Super
Yang-Mills Theories in Harmonic Superspace}
\end{center}
\vspace{5mm}

\begin{center}
I. L. Buchbinder$\,^{a)}$, E. I. Buchbinder$\,^{b)}$, 
S. M. Kuzenko\footnote{Alexander von Humboldt Research
Fellow. On leave from Department of Quantum Field Theory,
Tomsk State University, Tomsk 634050, Russia.}$\,^{, c)}$, and B. A.
Ovrut$\,^{d)}$
\end{center}

\begin{itemize}
\item[${}^{a)}$] \footnotesize{{\it Department of Theoretical Physics,
Tomsk State Pedagogical University, Tomsk 634041, Russia}}

\item[${}^{b)}$] \footnotesize{{\it Department of Quantum Field Theory,
Tomsk State University, Tomsk 634050, Russia }}

\item[${}^{c)}$] \footnotesize{{\it Institut f\"ur Theoretische Physik,
Universit\"{a}t Hannover, Appelstr. 2, 30167 Hannover, Germany}}

\item[${}^{d)}$] \footnotesize{{\it School of Natural Sciences,
Institute for Advanced Study, Olden Lane, Princeton, NJ 08540, USA\\
Department of Physics, University of
Pennsylvania, Philadelphia, PA 19104-6396, USA}}
\end{itemize}

\vspace{1cm}
\begin{abstract}
The background field method for $N=2$ super Yang-Mills theories in harmonic
superspace is developed. The ghost structure of the theory is
investigated. It is shown that the ghosts include two fermionic real
$\omega$-hypermultiplets (Faddeev-Popov ghosts) and one bosonic 
real $\omega$-hypermultiplet (Nielsen-Kallosh ghost), all in the adjoint
representation of the gauge group. The one-loop effective action is 
analysed in detail and it is found that its structure is determined only by
the ghost corrections in the pure super Yang-Mills theory. 
As applied to the case of $N=4$ super Yang-Mills theory, 
realized in terms of $N=2$ superfields, the latter result
leads to the remarkable conclusion that the one-loop effective action of the 
theory does not contain quantum corrections depending  on the $N=2$ gauge
superfield only. We show that the leading low-energy contribution to
the one-loop effective action in the $N=2$ $SU(2)$ super Yang-Mills theory
coincides with Seiberg's perturbative holomorphic effective action. 
\end{abstract}
\vfill

\end{titlepage}

\newpage
\setcounter{page}{1}
The background field method is a powerful and convenient tool for studying
the structure of quantum gauge theories. Its main idea is based on  
the so-called background-quantum splitting 
of the initial gauge fields into two parts:
the background fields and the quantum fields. To quantize the theory, one
imposes the gauge fixing conditions only on the quantum fields, introduces
the corresponding ghosts and considers the background fields as the
functional arguments of the effective action. The gauge fixing
functions are chosen to be background field dependent. As a result, we
can find in concrete gauge models a class of gauge fixing
functions with the property that the effective action will be invariant 
under the initial
gauge transformations. The background field method was originally suggested by
De Witt \cite{W1,W2} and then developed, and applied to concrete
theories, by a large number of authors. 
The attractive feature of the background field method is 
that it preserves the manifest gauge invariance at each step of
the loop calculations in quantum gauge theories.

Formulation of the background field method in $N=1$ super
Yang-Mills theory has been given in ref. \cite{GSR} and its
applications and generalizations were developed in detail (see 
\cite{GS}--\cite{GMZ} and also \cite{GGRS}--\cite{BK}). It turned
out that the background-quantum splitting in $N=1$ superfield Yang-Mills 
theory and supergravity is a non-trivial procedure as compared with the
conventional Yang-Mills and gravity theories.

Construction of the background field method in extended supersymmetric 
gauge theories faces a fundamental problem. The most 
natural and proper description of such theories should be formulated 
in terms of a suitable superspace and unconstrained superfields over it. 
Therefore, the first step to developing the background field method in 
extended supersymmetric theories is a solution of the problem of 
formulating these theories in terms of unconstrained superfields.

An approach to constructing the background field method for $N=2$ super
Yang-Mills theories in the standard $N=2$ superspace has been developed 
in ref. \cite{HST}. Some applications of this approach were investigated in
refs. \cite{M}. However, in our opinion, the approach 
of these authors looks very complicated from the technical point of view 
and its use for concrete problems should lead to a number of 
computational obstacles.

Interest in the quantum aspects of $N=2$ super Yang-Mills theories has
recently been revived by the seminal papers of 
Seiberg and Witten \cite{SW} (see \cite{B} for a review), 
where the non-perturbative
contribution to the low-energy effective action has been calculated. These
calculations were based on the general structure of the low-energy effective
action found in ref. \cite{S} (see also \cite{G}). The problem of the
effective action in the $N=2$ super Yang-Mills theory with matter has
recently been studied in refs. \cite{WGR}--\cite{LGRU}.  However, all these
computations of the effective action in $N=2$ super Yang-Mills theories 
were given in terms of $N=1$ superfields without manifest realization of the
$N=2$ supersymmetry.

The aim of this paper is to construct the background field method for $N=2$
super Yang-Mills theories and investigate the problem of the effective action
in terms of unconstrained $N=2$ superfields\footnote{We were 
informed by E. Ivanov that some aspects of the background field 
formulation for the $N=2$ super Yang-Mills theories were considered
by A. Galperin, E. Ivanov and E. Sokatchev in unpublished work
(private communication).}. We consider the formulation
of $N=2$ super Yang-Mills theory in the harmonic superspace approach
\cite{GIKOS}--\cite{Z}. This approach provides a clear understanding of
extended supersymmetric theories and opens opportunities to
investigate both classical and quantum aspects of such theories. As we
will see, the background field method formulation of $N=2$ super
Yang-Mills theory in harmonic superspace is relatively simple. In
particular, the structure of background-quantum splitting here is
much more similar to the conventional Yang-Mills theory than to 
the $N=1$ super Yang-Mills case.

We start with a brief review of the pure $N=2$ super Yang-Mills (SYM)
theory. In standard $N=2$ superspace with coordinates 
$z^M\equiv(x^m,\theta_i^\alpha, {\bar\theta}^i_{\dot\alpha})$, 
the gauge invariant action reads \cite{GSW}
\be
S_{{\rm SYM}}=\frac{1}{2g^2} {\rm tr}\int d^4xd^4\theta\, W^2=
\frac{1}{2g^2}{\rm tr} \int d^4xd^4{\bar \theta}\, {\bar W}^2
\label{1}
\ee
where $W$ and ${\bar W}$ are the covariantly chiral superfield strength
and its conjugate. These strengths are associated with the
gauge covariant derivatives
\be
{\cal D}_M\equiv ({\cal D}_m,{\cal
D}^i_\alpha,{\bar{\cal D}}_i^{\dot\alpha})= D_M+{\rm i}A_M \qquad
A_M=A_M^a(z)T^a
\label{2}
\ee
satisfying the algebra \cite{GSW}
\bea
& \{{\cal D}^i_\alpha,{\bar{\cal D}}_{{\dot\alpha j}}\}= -2{\rm i}\delta^i_j
{\cal D}_{\alpha{\dot\alpha}} \non \\
&\{{\cal D}^i_\alpha,{\cal D}^j_\beta\}=2{\rm i}\varepsilon_{\alpha\beta}
\varepsilon^{ij}{\bar W}\qquad \{{\bar{\cal D}}_{{\dot\alpha}i},
{\bar{\cal D}}_{{\dot\beta}j}\}=2{\rm i}\varepsilon_{{\dot\alpha}{\dot\beta}}
\varepsilon_{ij}W \\
&{[}{\cal D}_{\alpha{\dot\alpha}},{\cal D}^j_\beta]=\varepsilon_
{\alpha\beta}{\bar{\cal D}}^i_{\dot\alpha}{\bar W} \qquad
[{\cal D}_{\alpha{\dot\alpha}},{\bar{\cal D}}_{{\dot\beta}i}]=
\varepsilon_{{\dot\alpha}{\dot\beta}}{\cal D}_{\alpha i}W \non \;.
\label{3}
\eea
Here $D_M\equiv(\partial_m,D^i_\alpha,{\bar D}^{\dot\alpha}_i)$ are the
flat covariant derivatives, $T^a$ are the generators of the gauge group and
${\rm tr}\,(T^aT^b)=\delta^{ab}\,$.

The covariant derivatives and a matter superfield multiplet $\varphi(z)$
transform as follows
\be
{\cal D}'_M=e^{{\rm i}\tau}{\cal D}_M e^{-{\rm i}\tau} 
\qquad \varphi'=e^{{\rm i}\tau}\varphi
\label{4}
\ee
under the gauge group. Here $\tau=\tau^a(z)T^a$ and
$\tau^a={\bar\tau}^a$ are unconstrained real parameters. The set of all
transformations (\ref{4}) is said to form the $\tau$-group.

To realize the $N=2$ SYM theory as a theory of unconstrained dynamical
superfields, we extend the original superspace coordinates by bosonic
ones $||u_i{}^\mp||\in SU(2)$. These bosonic coordinates parametrize 
the two-sphere
$SU(2)/U(1)$ and extend the superspace to $N=2$ harmonic superspace
\cite{GIKOS}. Introducing the harmonic derivatives \cite{GIKOS}
\bea
D^{\pm\pm}=u^{\pm i}\frac{\partial}{\partial u^{\mp i}} &\qquad &
D^0=u^{+i}\frac{\partial}{\partial u^{+i}}-u^{-i}
\frac{\partial}{\partial u^{-i}} \non \\
{[}D^0,D^{\pm\pm}]=\pm 2D^{\pm\pm} & \qquad & [D^{++},D^{--}]=D^0
\label{5}
\eea
and defining
\be
{\cal D}_{\underline{M}}\equiv({\cal D}_M,{\cal D}^{++},{\cal D}^{--},
{\cal D}^0) \qquad {\cal D}^{\pm\pm}=D^{\pm\pm} \qquad
{\cal D}^0=D^0
\label{6}
\ee
one observes that the operators ${\cal D}_{\underline{M}}$ possess the
same transformation law (\ref{4}) with respect to the $\tau$-group as
${\cal D}_M$. If we now introduce
\be
{\cal D}^\pm_\alpha=u^\pm_i{\cal
D}^i_\alpha \qquad {\bar{\cal D}}^\pm_{\dot\alpha}=u^\pm_i{\bar{\cal
D}}^i_{\dot\alpha}
\label{7}
\ee
then the algebra of covariant derivatives takes the form
\bea
& \{{\cal D}^+_\alpha,{\cal
D}^+_\beta\}=\{{\bar{\cal D}}^+_ {\dot\alpha}, {\bar{\cal
D}}^+_{\dot\beta}\}= \{{\cal D}^+_\alpha,{\bar{\cal D}}^+
_\alpha\}=0\non \\
& \{{\cal D}^+_\alpha,{\cal D}^-_\beta\}=-2{\rm i}\varepsilon_{\alpha\beta}
{\bar W} \qquad
\{{\bar{\cal D}}^+_{\dot\alpha},{\bar{\cal D}}^-_{\dot\beta}\}=2{\rm i}
\varepsilon_{{\dot\alpha}{\dot\beta}}W \non \\
& \{{\bar{\cal D}}^+_{\dot\alpha},{\cal D}^-_\alpha\}=-\{{\cal D}^+_
\alpha,{\bar{\cal D}}^-_{\dot\alpha}\}=2{\rm i}{\cal D}_{\alpha{\dot\alpha}}
\non \\
& {[}{\cal D}^{++},{\cal D}^+_\alpha]=[{\cal D}^{++}{\bar{\cal D}}^+
_{\dot\alpha}]=0 \non \\
& {[}{\cal D}^{++},{\cal D}^-_\alpha]={\cal D}^+_\alpha \qquad
[{\cal D}^{++},{\bar{\cal D}}^-_{\dot\alpha}]={\bar{\cal D}}^+_{\dot
\alpha} \;.
\label{8}
\eea
The relations in the first line imply
\be
{\cal D}^+_\alpha=e^{-{\rm i}\Omega}D^+_\alpha e^{{\rm i}\Omega} \qquad
{\bar{\cal D}}^+_{\dot\alpha}=e^{-{\rm i}\Omega}{\bar D}^+_{\dot\alpha}
e^{{\rm i}\Omega}
\label{9}
\ee
for some Lie-algebra valued superfield $\Omega=\Omega^a(z,u)T^a$ with
zero $U(1)$-charge, $D^0\Omega^a=0$, and real, 
$\stackrel{\smile}{\Omega}{}^a = \Omega^a$, 
with respect to the analyticity-preserving conjugation \cite{GIKOS}
which we denote here by $\smile$. 
They allow one to define covariantly
analytic superfields constrained by
\be
{\cal D}^+_\alpha\Phi^{(q)}={\bar{\cal D}}^+_{\dot\alpha} \Phi^{(q)}=0\;.
\label{10}
\ee
Here $\Phi^{(q)}(z,u)$ carries $U(1)$-charge $q$, $D^0\Phi^{(q)}=q\Phi^
{(q)}$, and can be represented as follows
\be
\Phi^{(q)}=e^{-{\rm i}\Omega}\phi^{(q)} \qquad
D^+_\alpha\phi^{(q)}={\bar D}^+_{\dot\alpha}\phi^{(q)}=0
\label{11}
\ee
with $\phi^{(q)}(\zeta_A, u)$ being 
an unconstrained superfield over 
an analytic subspace of the harmonic superspace \cite{GIKOS} parametrized by 
$\zeta_A\equiv\{x^m_A,\theta^{+\alpha},{\bar\theta}^+_{\dot\alpha}\}$ 
and $u^\pm_i$, where $x^m_A = x^m - 2{\rm i}\theta^{(i}\sigma^m
{\bar \theta}^{j)}u^+_iu^-_j$ and
$\theta^\pm_\alpha=u^\pm_i\theta^i_\alpha$, ${\bar\theta}^\pm_{\dot\alpha}
=u^\pm_i{\bar \theta}^i_{\dot\alpha}$.

The $\Omega$ possesses a richer gauge freedom than the original $\tau$-group.
Its transformation law reads
\be
e^{{\rm i}\Omega'}=e^{{\rm i}\lambda}e^{{\rm i}\Omega}e^{-{\rm i}\tau}
\label{12}
\ee
with an unconstrained analytic gauge parameter $\lambda=\lambda^a(\zeta_A,u)
T^a$ being real with respect to the analyticity-preserving conjugation,
$\stackrel{\smile}{\lambda}{}^a =\lambda^a$.
The set of all
$\lambda$-transformations form the so-called $\lambda$-group
\cite{GIKOS}. The $\tau$-group acts on $\Phi^{(q)}$ and leaves
$\phi^{(q)}$ unchanged while the $\lambda$-group acts only on
$\varphi^{(q)}$ as follows
\be
\phi'^{(q)}=e^{{\rm i}\lambda}\phi^{(q)}\;.
\label{13}
\ee
The superfields $\Phi^{(q)}$ and $\varphi^{(q)}$ are said to correspond
to $\tau$- and $\lambda$-frames respectively. 

In the $\lambda$-frame, the covariant derivatives look like
\be
\nabla_{\underline{M}}=e^{{\rm i}\Omega}{\cal D}_{\underline{M}}
e^{-{\rm i}\Omega}
\label{14}
\ee
In particular
\bea
& \nabla^+_\alpha=D^+_\alpha \qquad{\bar\nabla}^+_{\dot\alpha}=
{\bar D}^+_{\dot\alpha}
\qquad \nabla^0=D^0 \non \\
& \nabla^{\pm\pm}=e^{{\rm i}\Omega}D^{\pm\pm}e^{-{\rm i}\Omega}
=D^{\pm\pm}+{\rm i}V^{\pm\pm}\;. 
\label{15}
\eea
In accordance with (\ref{8}), the connection $V^{++}=V^{++a}T^a$ is 
a real analytic superfield, $\stackrel{\smile}{V}{}^{++a}=V^{++a}$,
$D^+_\alpha V^{++}={\bar D}^+_{\dot\alpha}V^{++}=0$, and its 
transformation law is
\be
V'^{++}=e^{{\rm i}\lambda}V^{++}e^{-{\rm i}\lambda}
-ie^{{\rm i}\lambda}D^{++}e^{-{\rm i}\lambda}\;.
\label{16}
\ee
The analytic superfield $V^{++}$ turns out to be the single unconstrained
prepotential of the
pure $N=2$ SYM theory and all other objects are expressed
in terms of it. In particular, action (1) can be rewritten via
$V^{++}$ as follows \cite{Z}
\be
S_{{\rm SYM}}=\frac{1}{g^2} {\rm tr}\,
\int d^{12}z\sum\limits_{n=2}^\infty\frac{(-{\rm i})^n}
{n}\int du_1 du_2\dots du_n\frac{V^{++}(z,u_1)V^{++}(z,u_2)
\dots V^{++}(z,u_n)}
{(u^+_1u^+_2)(u^+_2u^+_3)\dots(u^+_nu^+_1)}\;.
\label{17}
\ee
The rules of integration over $SU(2)$ as well as the properties
of harmonic distributions are given in refs.
\cite{GIKOS,GIOS1}.

To quantize the theory under consideration we split $V^{++}$ into
{\it background} $V^{++}$ and {\it quantum} $v^{++}$ parts 
\be
V^{++}\rightarrow V^{++}+gv^{++}\;.
\label{18}
\ee
Then, the original infinitesimal gauge transformations (\ref{16}) can be
realized in two different ways:

(i) {\it background transformations}
\be
\delta V^{++}=-D^{++}\lambda -{\rm i}[V^{++},\lambda]=-\nabla^{++}\lambda
\qquad \delta v^{++}={\rm i}[\lambda,v^{++}]
\label{19}
\ee

(ii) {\it quantum transformations}
\be
\delta V^{++}=0 \qquad \delta v^{++}=-\frac{1}{g}\nabla^{++}\lambda 
-{\rm i}[v^{++},\lambda]\;.
\label{20}
\ee
It is worth pointing out that the form of the
background-quantum splitting (\ref{18}) and
the corresponding background and quantum transformations
(\ref{19}), (\ref{20}) are much more analogous to the conventional Yang-Mills
theory than to the $N=1$ non-abelian SYM model.
Our aim now is to construct an effective action as a gauge-invariant
functional of the background superfield $V^{++}$. 

Upon the splitting (\ref{18}), the classical action (\ref{17}) can be
shown to be given by
\bea
S_{{\rm SYM}}[V^{++}+gv^{++}]&=&S_{{\rm SYM}}[V^{++}] +
\frac{1}{4g}{\rm tr}\,\int d\zeta^{(-4)}
du\,v^{++}{\bar D}^+_{\dot\alpha}
{\bar D}^{+{\dot\alpha}}{\bar W}_\lambda \non \\
&+&\Delta S_{{\rm SYM}}[v^{++},V^{++}] \;.
\label{21}
\eea
Here 
$d\zeta^{(-4)}= d^4x_A d^2\theta^+
d^2{\bar\theta}^+$ and
\bea
\Delta S_{{\rm SYM}}[v^{++},V^{++}]&=&-{\rm tr}\,\int
d^{12}z\sum\limits_{n=2}^\infty {\frac{(-ig)}{n}}^{n-2}
\int du_1 du_2\dots du_n \non \\
& \times & \frac{v_\tau^{++}(z,u_1)v_\tau^{++}(z,u_2)\dots v_\tau^{++}(z,u_n)}
{(u^+_1u^+_2)(u^+_2u^+_3)\dots(u^+_nu^+_1)}
\label{22}
\eea
$W_\lambda$, ${\bar W}_\lambda$ and $v_\tau^{++}$ denote the $\lambda$- and
$\tau$-frame forms of $W$, $\bar W$ and $v^{++}$ respectively
\bea
& W_\lambda =e^{{\rm i}\Omega}W e^{-{\rm i}\Omega} \qquad
{\bar W}_\lambda =e^{{\rm i}\Omega}{\bar W} e^{-{\rm i}\Omega} \non \\
& v_\tau^{++}=e^{-{\rm i}\Omega}v^{++}e^{{\rm i}\Omega} \;.
\label{23}
\eea
The superfield $\Omega$ corresponds to the background covariant
derivatives constructed on the base of the
background connection $V^{++}$. The quantum action  
$\Delta S_{{\rm SYM}}$ given in (\ref{22}) depends on $V^{++}$ via the
dependence of $v^{++}_\tau$ on $\Omega$, the latter being a complicated
function of $V^{++}$. Each term in the action (\ref{21})
is manifestly invariant with respect to the background gauge transformations. 
The term linear in $v^{++}_\tau$ in (\ref{21}) determines the
equations of motion. This term should be dropped when considering the
effective action\footnote{As is well known, for calculating the effective
action within the loop expansion one really uses the construction $\Delta
S[\Psi,\psi]=S[\Psi+\psi]-S[\Psi]-S'[\Psi]\psi\,$, where the linear
term is absent (see, f.e., \cite{BOS}). Here $\Psi$ denotes the set of all
fields of the theory and we split $\Psi\rightarrow\Psi+\psi$, with
$\Psi$ the background field and $\psi$ the quantum one.}. 

To construct the effective action, we will follow the Faddeev-Popov
Ansatz. Within the framework of the background field method, 
we should fix only the quantum transformations (\ref{20}). 
Let us introduce the gauge fixing function in the form
\be
{\cal F}^{(4)}_\tau = D^{++}v^{++}_\tau=
e^{-{\rm i}\Omega}(\nabla^{++}v^{++})e^{{\rm i}\Omega}
= e^{-{\rm i}\Omega}{\cal F}^{(4)} e^{{\rm i}\Omega}
\label{24}
\ee
which changes by the law
\be
\delta {\cal F}^{(4)}_\tau = \frac{1}{g}e^{-{\rm i}\Omega}\{\nabla^{++}
(\nabla^{++}\lambda + {\rm i}g[v^{++},\lambda])\}e^{{\rm i}\Omega}
\label{25}
\ee
under the quantum transformations (\ref{20}). Eq. (\ref{25}) leads to
the Faddeev-Popov determinant
\be
\Delta_{{\rm FP}}[v^{++},V^{++}]={\rm Det}\,
\nabla^{++}(\nabla^{++}+igv^{++})\;.
\label{26}
\ee
To get a path-integral representation for $\Delta_{{\rm
FP}}[v^{++},V^{++}]$, 
we introduce two real analytic fermionic ghosts ${\bf b}$ and 
${\bf c}$, in the adjoint representation of the gauge group,
and the corresponding ghost action
\be
S_{{\rm FP}}[{\bf b},{\bf c},v^{++},V^{++}]=
{\rm tr} \, \int d\zeta^{(-4)}_A du\,
{\bf b}\nabla^{++}(\nabla ^{++}{\bf c}+{\rm i}g\,[v^{++},{\bf c}])\;.
\label{27}
\ee
As a result, we arrive at the effective action
$\Gamma_{{\rm SYM}}[V^{++}]$ in the form
\be
e^{{\rm i}\Gamma_{{\rm SYM}}[V^{++}]}=e^{{\rm i}S_{{\rm SYM}}[V^{++}]}
\int{\cal D}v^{++}{\cal D}{\bf b}
{\cal D}{\bf c}\,e^{{\rm i}(\Delta S_{{\rm SYM}}[v^{++},V^{++}]+
S_{{\rm FP}}[{\bf b},{\bf c},
v^{++},V^{++}])}\delta[{\cal F}^{(4)}-f^{(4)}]
\label{28}
\ee
where $f^{(4)}(\zeta_A,u)$ is an external Lie-algebra valued
analytic superfield independent of $V^{++}$, and $\delta[{\cal F}^{(4)}]$
is the proper functional analytic delta-function.

To transform the path integral for $\Gamma_{{\rm SYM}}[V^{++}]$ to a
more useful form, we average the right hand side in eq. (\ref{28}) with
the weight
\be
\Delta[V^{++}]\,\exp\left\{\frac{{\rm i}}{2\alpha}
{\rm tr}\,\int d^{12}zdu_1du_2\,
\,f_\tau^{(4)}(z,u_1)\frac{(u^-_1u^-_2)}{(u^+_1u^+_2)^3}f^{(4)}
_\tau(z,u_2)\right\}
\label{29}
\ee
Here $\alpha$ is an arbitrary (gauge) parameter. In flat superspace,
a weight function of this form has been used in refs. \cite{GIOS1,GIOS2}. The
functional $\Delta[V^{++}]$ should be chosen from the condition
\be
1=\Delta[V^{++}]\int{\cal D}f^{(4)}\exp\left\{\frac{{\rm i}}{2\alpha}
{\rm tr}\,\int
d^{12}z du_1du_2\,f^{(4)}_\tau(z,u_1)\frac{(u^-_1u^-_2)}{(u^+_1u^
+_2)^3}f^{(4)}_\tau(z,u_2)\right\}
\label{30}
\ee
hence
\bea
\Delta^{-1}[V^{++}]&=&\int{\cal D}f^{(4)}\,\exp\left\{\frac{{\rm i}}{2\alpha}
{\rm tr}\,\int d\zeta^{(-4)}_1d\zeta^{(-4)}_2du_1du_2\,f^{(4)}
(\zeta_1,u_1)A(1,2)f^{(4)}(\zeta_2,u_2)\right\}\non \\
&=&{\rm Det}^{-1/2}A
\label{31}
\eea
for a special background-dependent operator $A$ acting on the space of
analytic superfields with values in the Lie algebra
of the gauge group. Thus
\be
\Delta[V^{++}]={\rm Det}^{1/2}A \;.
\label{32}
\ee

To find ${\rm Det}\,A$ we represent it by a functional integral over
analytic superfields of the form
\be
{\rm Det}^{-1}A=\int{\cal D} \chi^{(4)}{\cal D}\rho^{(4)}
\exp \,\left\{ {\rm i}\;{\rm tr} \int d\zeta^{(-4)}_1du_1d\zeta^{(-4)}_2du_2\,
\chi^{(4)}(1)A(1,2)\rho^{(4)}(2) \right\}
\label{33}
\ee
and perform the following replacement of functional variables
\be
{}\rho^{(4)}=(\nabla^{++})^2\sigma \qquad{\rm Det}\left(
\frac{\delta\rho^{(4)}}
{\delta\sigma}\right) = {\rm Det}\,(\nabla^{++})^2\;.
\label{34}
\ee
Then we have
\bea
&{}&{\rm tr}\, \int d\zeta^{(-4)}_1du_1d\zeta^{(-4)}_2du_2\,\chi^{(4)}(1)
A(1,2)\rho^{(4)}(2) \non \\
&{}&={\rm tr}\, \int d^{12}zdu_1du_2\,\chi^{(4)}_\tau\frac{(u^-_1u^-_2)}
{(u^+_1u^+_2)^3}(D^{++}_2)^2\sigma_\tau(2)
=\frac{1}{2}{\rm tr}\int d^{12}z du\, 
\chi^{(4)}_\tau(D^{--})^2\sigma_\tau \non \\
&{}&=-{\rm tr}\,
\int d\zeta^{(-4)}du\,\chi^{(4)}\stackrel{\frown}{\Box}\sigma
\label{35}
\eea
where\footnote{
We use the notation 
$(D^+)^4 = \frac{1}{16} (D^+)^2 ({\bar D}^+)^2$, 
$(D^\pm)^2=D^{\pm \alpha} D^\pm_\alpha$, 
$({\bar D}^\pm)^2 = {\bar D}^\pm_{\dot{\alpha}}{\bar D}^{\pm \dot{\alpha}}$ 
and similar notation for the gauge-covariant derivatives.} 
\be
\stackrel{\frown}{\Box}=-\frac{1}{2}(\nabla^+)^4(\nabla^{--})^2
= -\frac{1}{2}(D^+)^4(\nabla^{--})^2\;.
\label{36}
\ee
On the basis of eqs. (\ref{32}--\ref{35}) one obtains
\be
\Delta[V^{++}]={\rm Det}^{-\frac{1}{2}}(\nabla^{++})^2
{\rm Det}^{\frac{1}{2}}\stackrel{\frown}{\Box}\;.
\label{36-1}
\ee
Below, it will be proven that
\be 
{\rm Det}\,\stackrel{\frown}{\Box} = 1\;.
\label{36-2}
\ee
Therefore, we are able to represent $\Delta[V^{++}]$ by the following
functional integral
\bea
\Delta[V^{++}]&=&
\int{\cal D}\phi\,
e^{{\rm i} S_{{\rm NK}}[\phi,V^{++}]} \non \\
S_{{\rm NK}}[\phi,V^{++}]  
&=& - \frac{1}{2} {\rm tr}\,\int d\zeta^{(-4)}du\,
\nabla^{++}\phi\nabla^{++}\phi
\label{37}
\eea
with the integration variable $\phi$ being a
bosonic real analytic superfield taking its values in the Lie algebra of the
gauge group. The $\phi$ is in fact the Nielsen-Kallosh ghost for the
theory.
As a result, we see that the $N=2$ SYM theory is described within the 
background field approach
by three ghosts: the two fermionic ghosts ${\bf b}$ and ${\bf c}$ and
the third bosonic ghost $\phi$. The ghost actions $S_{{\rm FP}}$ and
$S_{{\rm NK}}$ given by eqs. (\ref{27}) and (\ref{37}) correspond to the known
$\omega$-hypermultiplet \cite{GIKOS}.

Upon averaging the effective action
$\Gamma_{{\rm SYM}}[V^{++}]$ with the weight (\ref{29}), one gets 
the following path integral representation
\be
e^{{\rm i}\Gamma_{{\rm SYM}}[V^{++}]} = e^{{\rm i}S_{{\rm SYM}}[V^{++}]}
\int{\cal D}v^{++}{\cal D}{\bf b}{\cal D}{\bf c}{\cal D}\phi\,
e^{{\rm i}S_{{\rm Q}}[v^{++},{\bf b},{\bf c},\phi,V^{++}]}
\label{39}
\ee
where
\bea
S_{{\rm Q}}[v^{++},{\bf b},{\bf c},\phi,V^{++}]&=&
\Delta S_{{\rm SYM}}[v^{++},V^{++}]+S_{{\rm GF}}[v^{++},V^{++}]\non \\
&+&S_{{\rm FP}}[{\bf b},{\bf c},v^{++},V^{++}]+
S_{{\rm NK}}[\phi,V^{++}]\;.
\label{40}
\eea
Here $S_{{\rm GF}}[v^{++},V^{++}]$ is the gauge fixing contribution to the
quantum action
\bea
S_{{\rm GF}}[v^{++},V^{++}]=\frac{1}{2\alpha}{\rm tr}\,
\int d^{12}zdu_1du_2\frac
{(u^-_1u^-_2)}{(u^+_1u^+_2)^3}(D^{++}_1v^{++}_\tau(1))(D^{++}_2v^{++}_
\tau(2))\non \\
=\frac{1}{2\alpha}{\rm tr}\,\int d^{12}zdu_1du_2\,\frac
{v^{++}_\tau(1)v^{++}_\tau(2)}{(u^+_1u^+_2)^2}-\frac{1}{4\alpha}{\rm tr}\,\int
d^{12}zdu\,v^{++}_\tau(D^{--})^2v^{++}_\tau
\label{41}
\eea

Let us consider the sum of the quadratic part in $v^{++}$ 
of $\Delta S_{{\rm SYM}}$ (22) and $S_{{\rm GF}}$ (\ref{41}). It has the form
\be
\frac{1}{2}(1+\frac{1}{\alpha}) {\rm tr}\, \int d^{12}zdu_1du_2\,\frac{
v^{++}_\tau(1)v^{++}_\tau(2)}{(u^+_1u^+_2)^2}
+\frac{1}{2\alpha}{\rm tr}\, \int
d\zeta^{(-4)}du\,v^{++}\stackrel{\frown}{\Box}v^{++}
\label{42}
\ee
where we have used eq. (\ref{36}).
To further simplify the computation, we set $\alpha=-1$. 
We can now write the final result for the effective
action $\Gamma_{{\rm SYM}}[V^{++}]$
\be
e^{{\rm i}\Gamma_{{\rm SYM}}[V^{++}]}=e^{{\rm i}S_{{\rm SYM}}[V^{++}]}
\int{\cal D}v^{++}{\cal D}{\bf b}
{\cal D}{\bf c}{\cal D}\phi\,e^{{\rm i}S_{{\rm Q}}
[v^{++},{\bf b},{\bf c},\phi,V^{++}]}
\label{43}
\ee
where action $S_{{\rm Q}}$ is as follows
\bea
&{}& S_{{\rm Q}}[v^{++},{\bf b},{\bf c},\phi,V^{++}]=
S_2[v^{++},{\rm b},{\bf c},\phi,V^{++}]+
S_{{\rm int}}[v^{++},{\bf b},{\bf c},V^{++}] 
\label{44a}\\
&{}& S_2[v^{++},{\bf b},{\bf c},\phi,V^{++}]=
-\frac{1}{2} {\rm tr}\,\int d\zeta^{(-4)}du\,
v^{++}\stackrel{\frown}{\Box}v^{++}+{\rm tr}\,\int d\zeta^{(-4)}du\, 
{\bf b}(\nabla^{++})^2{\bf c} \non \\
&{}& \qquad \qquad \qquad \qquad \qquad + \frac{1}{2}{\rm tr}\, 
\int d\zeta^{(-4)}du\,\phi(\nabla^{++})^2\phi
\label{44b} \\
&{}& S_{{\rm int}}[v^{++},{\bf b},{\rm c},V^{++}]=
-{\rm tr}\,\int d^{12}zdu_1\dots du_n 
\sum\limits_{n=3}^\infty 
{\frac{(-ig)}{n}}^{n-2}\frac{v^{++}_\tau(z,u_1)\dots
v^{++}_\tau(z,u_n)}{(u^+_1u^+_2)\dots(u^+_nu^+_1)} \non\\
&{}& \qquad \qquad \qquad \qquad \qquad -{\rm i}g\, {\rm tr} \,\int 
d\zeta^{(-4)}du \nabla^{++}{\bf b}\;[v^{++}, {\bf c}]\;.
\label{44c}
\eea
Eqs. (\ref{43}--\ref{44c}) completely determine the structure of the
perturbation expansion for calculating the effective action
$\Gamma_{{\rm SYM}}[V^{++}]$ of the pure $N=2$ SYM theory 
in a manifestly supersymmetric
and gauge invariant form.

Let us now prove the relation (\ref{36-2}). We proceed by pointing out
that $\stackrel{\frown}{\Box}$ transforms each covariantly analytic superfield into 
a covariantly analytic one. When acting on spaces of such 
superfields, $\stackrel{\frown}{\Box}$ is equivalent to the second-order differential
operator
\bea
\stackrel{\frown}{\Box}_\tau=
e^{-{\rm i}\Omega}\stackrel{\frown}{\Box}e^{{\rm i}\Omega}  
&=&{\cal D}^m{\cal D}_m+
\frac{{\rm i}}{2}({\cal D}^{+\alpha}W){\cal D}^-_\alpha+\frac{{\rm i}}{2}
({\bar{\cal D}}^+_{\dot\alpha}{\bar W}){\bar{\cal D}}^{-{\dot\alpha}}-
\frac{{\rm i}}{4}({\bar{\cal D}}^+_{\dot\alpha}{\bar{\cal D}}^{+{\dot\alpha}}
{\bar W}){\cal D}^{--}\non \\
&+&\frac{{\rm i}}{4}({\cal D}^{-\alpha}{\cal D}^+_\alpha
W)+{\bar W}W 
\label{46}
\eea
as a consequence of the covariant derivative algebra (\ref{8}). It is 
remarkable that the differential part of $\stackrel{\frown}{\Box}$ is uniquely
determined from the requirements
that (i) $\stackrel{\frown}{\Box}$ 
is constructed in terms of the covariant derivatives
only; (ii) $\stackrel{\frown}{\Box}$ 
moves every covariantly analytic superfield into
a covariantly analytic one; (iii) $\stackrel{\frown}{\Box}$ 
is a second-order operator
containing the only term ${\cal D}^m {\cal D}_m$ with two vector covariant
derivatives. 
Next, we introduce the  proper-time
representation for a regularized form of ${\rm
Det}\,\stackrel{\frown}{\Box}$
(see ref. \cite{BK} for more details
of the superfield proper-time technique)
\be
\ln \,({\rm Det}\,\stackrel{\frown}{\Box})_{\rm reg}=
-\mu^{2\varepsilon}\int\limits_0
^\infty d({\rm i}s)({\rm i}s)^{\varepsilon-1}{\rm Tr}\,
e^{{\rm i}s\stackrel{\frown}{\Box}}.
\label{47}
\ee
Here we have introduced the regularization parameters $\mu$ and
$\varepsilon$, where $\varepsilon\rightarrow 0$ in the end of calculations.
The ${\rm Tr}$ of the analytic `heat kernel'
$e^{{\rm i}s\stackrel{\frown}{\Box}}$ is
defined by
\be
{\rm Tr}\,e^{{\rm i}s\stackrel{\frown}{\Box}}=
{\rm tr}\,\int d\zeta^{(-4)}_1du_1
d\zeta^{(-4)}_2du_2\,
\delta^{(2,2)}_A(\zeta_1,u_1|\zeta_2,u_2)e^{{\rm i}s\stackrel{\frown}{\Box}}
\delta^{(2,2)}_A(\zeta_1,u_1|\zeta_2,u_2)
\label{48}
\ee
and $\delta^{(2,2)}_A (1,2)$ denotes the proper analytic subspace
delta-function
\bea
\delta^{(2,2)}(\zeta_1,u_1|\zeta_2,u_2)&=&(D^+_1)^4\{\delta^{12}(z_1-z_2)
\delta^{(-2,2)}(u_1,u_2)\} \non \\             
&=&(D^+_2)^4\{\delta^{12}(z_1-z_2)
\delta^{(2,-2)}(u_1,u_2)\}
\label{49}
\eea
with $\delta^{(-2,2)}(u_1,u_2)$ and $\delta^{(2,-2)}(u_1,u_2)$ being
special harmonic delta-functions \cite{GIOS1}. 
Further, we rewrite
\be
e^{{\rm i}s\stackrel{\frown}{\Box}} =
1 + \stackrel{\frown}{\Box}
\frac{e^{{\rm i}s\stackrel{\frown}{\Box}}-1}{\stackrel{\frown}{\Box}}
= 1-\frac{1}{2}(D^+)^4(\nabla^{--})^2 \cU (s)
\label{iv3}
\ee
where
\be
\cU (s) =
\frac{e^{{\rm i}s\stackrel{\frown}{\Box}}-1}{\stackrel{\frown}{\Box}}\;.
\label{iv4}
\ee
Then
\bea
{\rm Tr}\,e^{{\rm i}s\stackrel{\frown}{\Box}}&=&
 -\frac{1}{2}
{\rm tr}\,\int d\zeta^{(-4)}_1du_1
d\zeta^{(-4)}_2du_2\,
\delta^{(2,2)}_A(\zeta_1,u_1|\zeta_2,u_2)\non \\
& \times &(D^+_1)^4 (\nabla^{--}_1)^2 \cU(s)
\delta^{(2,2)}_A(\zeta_1,u_1|\zeta_2,u_2)\;.
\label{iv7}
\eea
Taking into account the explicit form of $\stackrel{\frown}{\Box}$
and making use of 
the covariant derivative algebra (\ref{8}), one readily observes
\bea
\cU_\t (s) &=& e^{-{\rm i}\Omega} \cU (s) e^{{\rm i}\Omega}\non \\
&=& A(s)+B^{+\alpha}(s){\cal D}^-_\alpha+{\tilde B}
^{+{\dot\alpha}}(s){\bar{\cal D}}^-_{\dot\alpha}+C^{++}(s)({\cal D}^-)
^2+{\tilde C}^{++}(s)({\bar{\cal D}}^-)^2 \non \\
&+&E^{++\alpha{\dot\alpha}}(s)[{\cal D}^-_\alpha,{\bar{\cal D}}^-_{\dot
\alpha}]+F^{(3)\alpha}{\cal D}^-_\alpha({\bar{\cal D}}^-)^2+{\tilde F}
^{(3){\dot\alpha}}(s){\bar{\cal D}}^-_{\dot\alpha}({\cal D}^-)^2 \non\\
&+&G^{(4)}(s)({\cal D}^-)^4\;.
\label{iv6}
\eea
Here $A$, $B^{+\alpha}$, ${\tilde B}^{+{\dot\alpha}}$, $C^{++}$,
${\tilde C}^{++}$, $E^{++\alpha{\dot\alpha}}$, $F^{(3)\alpha}$,
${\tilde F}^{(3){\dot\alpha}}$ and $G^{(4)}$ are some functions of
the real parameter $s$, the covariant derivatives
${\cal D}_m$, ${\cal D}^{--}$ as well as of $W$, ${\bar W}$ and their
covariant derivatives. The exact form  of these functions is not
essential here. The dependence of $\cU_\t (s)$
on the spinor
covariant derivatives has been written down in eq. (\ref{iv6}) explicitly
and this is all that we will need later on.

The integrals over the analytic subspace in eq. (\ref{iv7}) can be transformed
into integrals over the full superspace
\bea
{\rm Tr}\,e^{{\rm i}s\stackrel{\frown}{\Box}}
&=& -\frac{1}{2}
{\rm tr}\,\int d^{12}z_1du_1d^{12}z_2du_2
\delta ^{12}(z_1-z_2)\delta^{(2,-2)}(u_1,u_2)\non\\
&\times &(\nabla^{--}_1)^2 \cU (s)
\delta^{(2,2)}_A(\zeta_1,u_1|\zeta_2,u_2)\non \\
&=& -\frac{1}{2}
{\rm tr}\,\int d^{12}z_1du_1d^{12}z_2du_2
\delta ^{12}(z_1-z_2)\delta^{(2,-2)}(u_1,u_2)\non\\
& \times & (D^{--}_1)^2 \cU_\t (s) (\cD^+_1)^4
\delta ^{12}(z_1-z_2)\delta^{(-2,2)}(u_1,u_2)\;.
\label{iv9}
\eea
Because of eq. (\ref{iv6}), we can continue in the manner
\bea
{\rm Tr}\,e^{{\rm i}s\stackrel{\frown}{\Box}}
&=& -\frac{1}{2}
{\rm tr}\,\int d^{12}z_1du_1d^{12}z_2du_2
\delta ^{12}(z_1-z_2)\delta^{(2,-2)}(u_1,u_2)\non\\
&\times &(D^{--}_1)^2 G^{(4)}(s) (\cD^-_1)^4 (\cD^+_1)^4
\delta ^{12}(z_1-z_2)\delta^{(-2,2)}(u_1,u_2) \non \\
&=& -\frac{1}{2}
{\rm tr}\,\int d^{12}z_1du_1d^{12}z_2du_2
[ \d^8(\q_1 - \q_2) (D^-_1)^4 (D^+_1)^4 \d^8(\q_1 - \q_2)]\non \\
& \times & \d^4 (x_1 - x_2) [(D^{--}_1)^2 \delta^{(2,-2)}(u_1,u_2)]
G^{(4)} (s) \delta^{(-2,2)}(u_1,u_2) \d^4 (x_1 - x_2)\;.
\label{iv10}
\eea
Since
$$
\int d \q^8_2 \,\d^8(\q_1 - \q_2) (D^-_1)^4 (D^+_1)^4 \d^8(\q_1 - \q_2)
= 1
$$
we obtain
\bea
{\rm Tr}\,e^{{\rm i}s\stackrel{\frown}{\Box}}
&=& -\frac{1}{2}
{\rm tr}\,\int d^8 \q d^{4}x_1du_1d^{4}x_2du_2
\d^4 (x_1 - x_2)\non \\
&\times & [(D^{--}_1)^2 \delta^{(2,-2)}(u_1,u_2)]
G^{(4)} (s) \delta^{(-2,2)}(u_1,u_2) \d^4 (x_1 - x_2) = 0
\label{iv11}
\eea
as a consequence of the following property of harmonic delta-functions
\bea
&&\int du_1 du_2 \,\d^{(m,-m)}(u_1,u_2)\, f^{(\pm 2p)}(u_1)\,
(D_1^{\mp\mp})^p\,
\d^{(-m,m)}(u_1,u_2) = 0 \quad \qquad p>0
\label{iv1}\\
&&\int du_1 du_2 \,\d^{(0,0)}(u_1,u_2)\, f^{(0)}(u_1)\,
\d^{(0,0)}(u_1,u_2) = f^{(0)}(0)\,\infty \qquad \qquad  \qquad p=0
\label{iva2}
\eea
with $f^{(\pm 2p)}(u)$ an arbitrary function
of $U(1)$-charge $\pm 2p$. Eqs. (\ref{iv1}) and (\ref{iva2}) can be 
readily justified if one regularizes the harmonic delta-function
\cite{GIOS1}
\be
\d^{(m,-m)}(u_1,u_2) = \sum_{n=0}^{\infty} (-1)^{n+m}
\frac{(2n + m + 1)!}{n!(n+m)!} (u^+_1)_{(n+m} (u^-_1)_{n)} (u^+_2)^{(n}
(u^-_2)^{n+m)}
\label{iva4}
\ee
by cutting off the Fourier series at the upper limit
\be
\d_N^{(m,-m)}(u_1,u_2) \equiv \sum_{n=0}^{N} (-1)^{n+m}
\frac{(2n + m + 1)!}{n!(n+m)!} (u^+_1)_{(n+m} (u^-_1)_{n)} (u^+_2)^{(n}
(u^-_2)^{n+m)}
\label{iva5}
\ee
where $N \rightarrow \infty$ in the end of the calculation. Eq. (\ref{iv1})
remains valid for the regularized harmonic delta-functions.
Therefore, we have proven relation (\ref{36-2}).

The generic expressions (\ref{43}--\ref{44c}) open an opportunity to
investigate the loop corrections to the effective action 
$\Gamma_{{\rm SYM}}[V^{++}]$.
Let us consider the one-loop approximation. In this case the effective
action has the structure
$\Gamma_{{\rm SYM}}[V^{++}]=S_{{\rm SYM}}[V^{++}]+
\Gamma_{{\rm SYM}}^{(1)}[V^{++}],$ where
$\Gamma_{{\rm SYM}}^{(1)}[V^{++}]$ describes the one-loop quantum corrections.
The relations (\ref{43}) and (\ref{44a}) along with eq. (\ref{36-2})
immediately lead to
\be
\Gamma_{{\rm SYM}}^{(1)}[V^{++}]=-{\rm i}\left(
{\rm Tr}\ln(\nabla^{++})^2-\frac{1}{2}
{\rm Tr}\ln(\nabla^{++})^2\right)=-
\frac{{\rm i}}{2}{\rm Tr}\ln
(\nabla^{++})^2\,.
\label{45}
\ee
It is remarkable that even if the relation (\ref{36-2}) was not true,
${\rm Tr} \ln \stackrel{\frown}{\Box}$ 
would not enter $\Gamma^{(1)}[V^{++}]$ anyway.
In such a case we should start from eq. (\ref{36-1}), keeping 
${\rm Det}\stackrel{\frown}{\Box}$ intact at all stages, 
and would obtain for
$\Gamma^{(1)}[V^{++}]$ the following representation
\be
\Gamma^{(1)}[V^{++}]=-{\rm i}\left(\frac{1}{2}
{\rm Tr}\ln \stackrel{\frown}{\Box} - \frac{1}{2}
{\rm Tr}\ln \stackrel{\frown}{\Box} \right)
-\frac{{\rm i}}{2}{\rm Tr}\ln
(\nabla^{++})^2\,.
\label{45-1}
\ee
As a result, the whole contribution to the 
one-loop effective action is stipulated only by the ghost contribution.
Moreover, this ghost contribution differs only in sign 
from the 
contribution of a single real $\omega$-hypermultiplet, 
in the adjoint representation of
the gauge group, coupled to the external gauge superfield $V^{++}$. 
The structure of the
effective action of the $\omega$-multiplet has been investigated in our
previous paper \cite{BBIKO} for an abelian gauge group, and that work
is readily extended to the non-abelian case.

We have developed the background field method for the pure $N=2$
SYM theory. In the general case, the classical action contains not only the 
pure SYM part given by (\ref{1}) (or, what is equvalent, by (\ref{17})), 
but also the matter action of the general form \cite{GIKOS}
\be 
S_{{\rm MAT}}=
- \int d\zeta^{(-4)}
du\stackrel{\smile}{q}{}^+ \nabla^{++}q^+ -
\frac{1}{2}\int d\zeta^{(-4)}
du\,\nabla^{++}\omega^{\rm T}\nabla^{++}\omega
\label{hyp}
\ee
describing the matter $q$-hypermultiplet $(q^+(\zeta_A,u),
\stackrel{\smile}{q}{}^+(\zeta_A,u))$ and $\omega$-hypermultiplet
$\omega (\zeta_A,u)$ coupled to the SYM gauge superfield $V^{++}$.
Our previous considerations can be easily extended to the case of the
general $N=2$ SYM theory. The only non-trivial new information, however, is the
explicit structure of the matter superpropagators associated with
the above action (\ref{hyp}). They read as follows
\bea
G_{\rm F}^{(1,1)}(1,2) &\equiv& 
{\rm i}\,<q^+(1)\,\stackrel{\smile}{q}{}^+(2)> \non \\
&=& - \frac{1}{\stackrel{\frown}{\Box}{}_1}
(D_1^+)^4(D_2^+)^4 \left\{\delta^4(x_1-x_2)\delta^8(\theta_1-\theta_2)
{1\over (u^+_1 u^+_2)^3}e^{{\rm i}\Omega (1)}e^{-{\rm i}\Omega (2)} \right\}
\label{qgreen}\\
G_{\rm F}^{(0,0)}(1,2) &\equiv& {\rm i}\,<\omega(1)\,\omega^{\rm T}(2)> 
\non \\
&=& -\frac{1}{\stackrel{\frown}{\Box}{}_1}
(D_1^+)^4(D_2^+)^4 \left\{\delta^4(x_1-x_2)\delta^8(\theta_1-\theta_2)
{(u^-_1 u^-_2)\over (u^+_1 u^+_2)^3}
e^{{\rm i}\Omega (1)}e^{-{\rm i}\Omega (2)}\right\}
\label{ogreen}
\eea
and satisfy the equations
\bea
\nabla_1^{++}G_{\rm F}^{(1,1)}(1,2)&=& \delta_A^{(3,1)}(1,2)\\
(\nabla_1^{++})^2G_{\rm F}^{(0,0)}(1,2)&=& - \delta_A^{(4,0)}(1,2)
\eea
respectively, with $\stackrel{\frown}{\Box}$ given by (\ref{46}). 
Switching off the gauge superfield, the Green's functions
turn into the free ones obtained in \cite{GIOS1}. The Green's functions
(\ref{qgreen}) and (\ref{ogreen}) are to be used for loop calculations
in the background field approach.

As the simplest application of the techniques developed here, we demonstrate 
the fact that the one-loop quantum correction to the effective action 
of the $N=4$ SYM theory realized in terms of $N=2$ superfields does
not contain contributions depending only on the $N=2$ gauge superfield.
In $N=2$ superspace, this theory is described by the action
\be
S^{N=4}_{{\rm SYM}}=\frac{1}{2g^2} {\rm tr}\int d^4xd^4\theta\, W^2
\,-\,\frac{1}{2g^2}\,{\rm tr}\,\int d\zeta^{(-4)}
du\,\nabla^{++}\omega\nabla^{++}\omega
\label{n4}
\ee
with $\omega$ the real $\omega$-hypermultiplet 
taking its values in the Lie algebra of the gauge group. This action was shown
to possess $N=4$ supersymmetry \cite{GIOS2} transforming $V^{++}$ and
$\omega$ into each other. We denote by $\Gamma[V^{++},\omega]$
the effective action of the theory and consider the one-loop
correction $\Gamma^{(1)}[V^{++},\omega]$. The contributions to
$\Gamma^{(1)}[V^{++},\omega]$, which depend only on $V^{++}$, come from
(\ref{45}) as well from the matter functional integral
\be
e^{{\rm i}\Gamma^{(1)}_{{\rm MAT}}[V^{++}]}= \int {\cal D}\omega
e^{-{\rm i}\frac{1}{2g^2}\,{\rm tr}\,\int d\zeta^{(-4)}
du\,\nabla^{++}\omega\nabla^{++}\omega} \qquad
\Gamma^{(1)}_{{\rm MAT}}[V^{++}] =
\frac{{\rm i}}{2}{\rm Tr}\ln
(\nabla^{++})^2\,.
\ee
But $\Gamma_{{\rm SYM}}^{(1)}[V^{++}]$ and 
$\Gamma^{(1)}_{{\rm MAT}}[V^{++}]$
exactly cancel each other.

Finally, we would like to discuss the leading low-energy contribution
to the one-loop effective action in the $N=2$ $SU(2)$ SYM theory with the
gauge group spontaneously broken to $U(1)$. Here the one-loop effective
$\Gamma^{(1)}_{SU(2)}[V^{++}]$ reads 
\be
\Gamma^{(1)}_{SU(2)} [V^{++}]= - \Gamma_\phi[V^{++}]
\label{64}
\ee
with $\Gamma_\phi[V^{++}]$ the effective action of a real
$\omega$-hypermultiplet 
in the adjoint representation of $SU(2)$ coupled to the external gauge 
superfield $V^{++}$ :
\be
e^{ {\rm i}\,\Gamma_\phi[V^{++}]} = \int {\cal D}\phi 
\exp \left\{ 
- \frac{{\rm i}}{2}\,{\rm tr}\, \int  d \zeta^{(-4)}  du 
\nabla^{++} \phi\,\nabla^{++} \phi \right\} 
\label{65}
\ee
where
\bea
& \phi = \phi^a \tau^a \qquad \nabla^{++}\phi = D^{++} \phi 
+ {\rm i}[V^{++},\phi]
\nonumber \\
&\tau^a =\frac{1}{\sqrt{2}} \sigma^a \qquad
[\tau^a, \tau^b] ={\rm i} \sqrt{2}\varepsilon^{abc} \tau^c \qquad
{\rm tr}\,(\tau^a \tau^b)= \delta^{ab}\;.
\label{66}
\eea

Upon the spontaneous breakdown of $SU(2)$, only the $U(1)$ gauge symmetry
survives and
the gauge superfield $V^{++} =V^{++a} \tau^a$ takes the form
\be
V^{++} = V^{++3} \tau^3 \equiv {\cal V}^{++}\tau^3\;.
\label{67}
\ee
Here ${\cal V}^{++}$ consists of two parts,
${\cal V}^{++} = {\cal V}_0^{++} + {\cal V}_1^{++}$, where 
${\cal V}_0^{++}$
corresponds to a constant strength ${\cal W}_0 = const$,
and ${\cal V}_1^{++}$ is an abelian gauge superfield.
It can be proved that the presence of ${\cal V}_0^{++}$ leads to the appearance
of mass $|{\cal W}_0|^2$ for matter multiplets (see \cite{BBIKO}). 
Now, we have 
\bea
\nabla^{++}\phi^1 &=& D^{++} \phi^1 + \sqrt{2}{\cal V}^{++}
\phi^2 \nonumber\\
\nabla^{++}\phi^2 &=& D^{++} \phi^2 - \sqrt{2}{\cal V}^{++}
\phi^1 \nonumber\\
\nabla^{++}\phi^3 &=& D^{++} \phi^3\;.
\label{68}
\eea
Thus $\phi^3$ completely decouples. Unifying $\phi^1$ and $\phi^2$
in to the {\it complex} $\omega$-hypermultiplet
$\omega = \phi^1 - {\rm i} \phi^2$,
we observe
\be
\nabla^{++}\omega = D^{++} \omega + {\rm i} \sqrt{2}{\cal V}^{++} \omega
\label{69}
\ee
hence the $U(1)$-charge of $\omega$ is $e= \sqrt{2}$. In our previous
paper \cite{BBIKO} it was shown that the effective actions of 
the charged complex $\omega$-hypermultiplet and the charged
$q$-hypermultiplet, interacting with background $U(1)$ gauge
superfield ${\cal V}^{++}$, are related by 
$\Gamma_\omega [{\cal V}^{++}]=2\Gamma_q [{\cal V}^{++}]$ and the leading
contribution to $\Gamma_q [{\cal V}^{++}]$ in the massive theory is given
by
\be
\Gamma_q [{\cal V}^{++}] = \int d^4 x d^4 \theta {\cal F}({\cal W}) + 
{\rm c.c.}
\qquad {\cal F}({\cal W})=- \frac{e^2}{64 \pi^2} {\cal W}^2 
\ln \frac{{\cal W}^2}{M^2}\,.
\label{70}
\ee
Here $e$ is the charge of $q^+$ (it coincides with the charge of $\omega$
in the above correspondence), $M$ is the renormalization scale, 
and ${\cal W}$ the 
chiral superfield strength associated with ${\cal V}^{++}$. 
Since in our
case $e=\sqrt{2}$, and taking into account eq. (\ref{64}), we finally obtain 
\be
\Gamma^{(1)}_{SU(2)} [{\cal V}^{++}] =\frac{1}{16 \pi^2}
\int d^4 x d^4 \theta\,{\cal W}^2 \ln \frac{{\cal W}^2}{M^2}\,.
\label{71}
\ee
This is exactly Seiberg's low-energy effective action \cite{S} 
found by integrating 
the $U(1)$ global anomaly and using the component analysis (note that
Seiberg used the strength $\Psi = \sqrt{2} {\cal W}$).
  
Let us summarize the results. We have considered $N=2$ super Yang-Mills
theories in harmonic superspace and formulated the background field
method for these theories. For the pure $N=2$ SYM theory, 
the effective action is given by a path integral
over the quantum gauge superfields and the fermionic and bosonic ghosts
corresponding to $\omega$-hypermultiplets. This path integral representation
allows one to
carry out the perturbative loop calculations in the theory under
consideration in a manifestly $N=2$ supersymmetric and gauge invariant
manner. The structure of the one-loop contributions to the effective action has
been investigated and it has been shown that the whole one-loop contribution is
stipulated only by the ghosts in the form of the effective action of the fermionic
$\omega$-hypermultiplet coupled to the external super Yang-Mills field. This
result has been applied to calculating the one-loop effective action in the
$N=4$ super Yang-Mills theory treated as the $N=2$ super
Yang-Mills theory coupled to the $\omega$-hypermultiplet in the 
adjoint representation of the gauge group. Taking into account the
structure of the one-loop effective action in the pure $N=2$ SYM theory, we
conclude that the one-loop effective action in the $N=4$ SYM theory 
does not contain corrections depending on the $N=2$ gauge superfield only.
Finally, we have derived the well-known Seiberg's low-energy effective
action in the harmonic superspace approach.

\vspace{1cm}
{\bf Acknowledgements}. We are grateful to E.A. Ivanov for collaboration
at the early stage of this work and for useful comments. 
I.L.B. and S.M.K. acknowledge the
partial support from the RFBR-DFG project No. 96-02-001800
and the RFBR project No. 96-02-16017. 
I.L.B. is very grateful to the
Department of Physics, University of Pennsylvania for hospitality
and for support from Research Foundation of the University of Pennsylvania. 
S.M.K. is grateful to the Alexander von Humboldt Foundation for support.
B.A.O. acknowledges the DOE Contract No. DE-AC02-76-ER-03072 
for partial support.

\end{document}